# On the Karman constant

**Trinh, Khanh Tuoc**

Institute Of Food Nutrition and Human Health

Massey University, New Zealand

*K.T.Trinh@massey.ac.nz*

**ABSTRACT**

Numerous studies in the past 40 years have established that turbulent flow fields are populated by transient coherent structures that represent patches of fluids moving cohesively for significant distances before they are worn out by momentum exchange with the surrounding fluid. Two particular well-documented structures are the hairpin vortices that move longitudinally above the wall and ejections inclined with respect to the wall that bring the fluid from the transient viscous layers underneath these vortices into the outer region of the boundary layer.

It is proposed that the Karman universal constant in the logarithmic law the sine of the angle between the transient ejections and the direction normal to the wall. The edge of the buffer layer is represented by a combination of the Karman constant and the damping function in the wall layer.

Computation of this angle from experimental data of velocity distributions in turbulent shear flows matches published traces of fronts of turbulence obtained from the time shifts in the peak of the correlation function of the velocity.

Key works: Turbulence, coherent structures, Karman's constant, mixing-length, shear layers

## 1 Introduction

The so-called 'log law' is widely used to describe most turbulent wall-bounded flows, and lies at the core of the most widely used engineering computational models involving turbulent flow near surfaces. The most common (but not the only) expression of this law is through the mean velocity profile normalized with so-called 'inner' variables given by

$$U^+ = A \ln y^+ + B = \frac{1}{\kappa} \ln y^+ + B \qquad (1)$$

where $U/u_*$, $y^+ = yu_*/\nu$, $u_* = \sqrt{\tau_w/\rho}$, $\tau_w$ is the wall shear stress, $\nu$ the kinematic viscosity and $\rho$ the density. Millikan (1938) derived this law by an overlap argument between an outer region scaling with the outer variables and a wall region scaling with wall variables. Prandtl (1935) derived the same form of the law from an estimate of a typical turbulence scale that he called the mixing-length. The parameter $\kappa$ is known as Karman's constant and was thought to be a universal constant with a value of 0.4. But this canonical value - nomenclature of Andreas (2009)- is not accepted by all authors.

For example, George (2007) concluded that $\kappa = 0.42$ for pipe flow from the work of McKeon et al (2004) on the Princeton superpipe data and $\kappa = 0.38$ for boundary layer flow from the work of Österlund (2000). There is far less agreement among measurements in atmospheric conditions with values as low as 0.35 (Businger et al., 1971) and as high as 0.46 (Sheppard, 1947). Högström (1985) obtained a value of 0.4 but Telford and Businger (1986) draw his analysis into question and underline the enormous methodical problems in connection with the evaluation of experimental atmospheric data. There have thus been attempts to evaluate the value of $\kappa$ by other methods than using equation (1). Lo et al. (2005) estimated $\kappa$ from homogeneous instead of wall bounded turbulence and obtained a value of 0.42. Orsag and Patera (1981) obtained a value 0.46 by numerical integration of the NS equations but caution about the limitations of their solution. Zhang et al (2008) used a variational method to obtain a value between 0.383 and 0.390. Vogel and Frenzen (2002) argued that $\kappa$ is a weak function of surface roughness but Högström et al. (2002) dispute this conclusion. From their calculation using two large atmospheric data sets, Andreas et al. (2006) suggest that is $\kappa$ a constant at 0.387±0.003 for a broad range of roughness Reynolds number and in his latest discussion Andreas (2009) propose the value of 0.39. In their review of simulations of wall turbulence, Jimenez and Moser (2007) settle on a value of 0.4.

Zhao et al. (2002) concluded from their pipe flow data that $\kappa$ increases slowly with the Reynolds number. Barenblatt (1997) has argued that a universal velocity would imply a constant value of $\kappa$ but well known data indicate that the normalised velocity profile is Reynolds number dependent. By dimensional analysis Barenblatt show that a power law sinilar profile is more indicated.

George (2007) noted that the differences $\kappa$ in may appear slight, especially given the errors involved in most turbulence measurements, but in fact they are quite important in practical applications. The reason for this is that most turbulence models used by the aircraft industry depend in one form or another on the assumption that the flow very close to the wall can be described by the logarithmic velocity profile of equation (1) and, by one estimate, a 2% decrease in k makes a 1% decrease in the overall drag for an aircraft. The constant $\kappa$ may also be used as a criterion for evaluating the correctness of theories of turbulence. George asks the question "Is there a universal log law for turbulent wall-bounded flows?"

There have been a number of attempts at a theoretical prediction of the Karman constant. For example Telford (1982) predicted $\kappa = 0.37$ using a formula which connects two semi-empirical parameters, the entrainment and the decay constants. Yakhot and Orsag (1986) predicted $\kappa = 0.372$ using re-normalisation group analysis of turbulence. Some authors propose that the Karman constant is purely a mathematical constant. For example Bergmann (1998) matched asymptotic profiles of a viscous flow and an inviscid turbulent flow to give $\kappa = \ln(e) = 0,379$ but his assumptions are disputed by Andreas and Trevino (2000). Baumert (2009) introduced a value $\kappa = 1/\sqrt{2\pi}$ as part of model comparing turbulent fields to an ensemble of vortex-dipole tubes and real gases of semi-stable macromolecules.

Despite the lack of consensus, McKeon and K. R. Sreenivasan (2007) noted that "It is heartening that one can, these days, have enough confidence in (laboratory) measurements to claim that 0.38 and 0.42 are, in fact, different without rounding them both off to a 'universal' value of 0.4."

This paper explores the significance of Karman's constant.

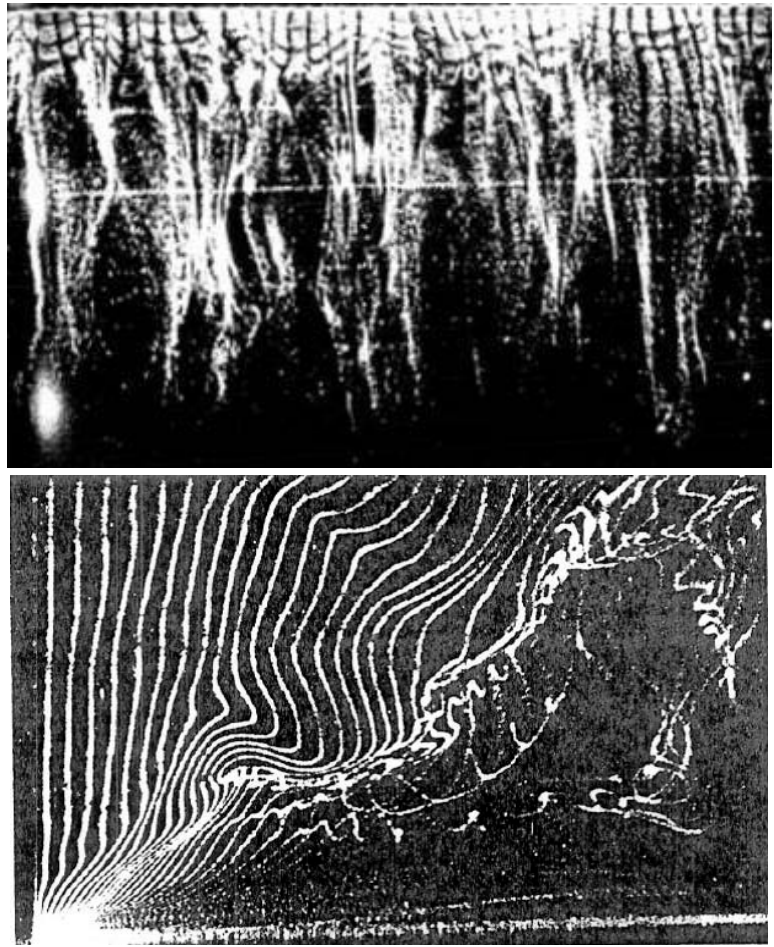

Figure 1 (Top) alternate low and high speed streaks (Kline et al. 1967). (Bottom) Sweep and burst at the wall (Offen and Kline 1971).

## 2 Background Theory

### 2.1 Coherent structures in turbulence

Modern turbulence research was triggered by the classic work of Kline et al. (1967) who showed by hydrogen bubble visualisation that the wall layer was not quiescent, as might be expected from the laminar sublayer postulate of Prandtl (1935). Kline et al. found that slow-speed fluid from the wall layer was periodically ejected into the main flow stream through a violent "bursting" process accompanied from an inrush of fast fluid from the outer flow. As the fast fluid rushed towards the wall, the streamline became contorted into a vortex pattern. The travelling vortex induces underneath its path (Figure 1) which is observed as streaks of low-speed fluid. The streaks tend to lift, oscillate and eventually burst in violent ejections from the wall towards the outer region. The process of turbulence production in the wall layer has been well documented (Kline et al., 1967, Kim et al., 1971, Offen and Kline, 1974, Corino and Brodkey, 1969).

Transient structures embedded in the flow field have been found not only near the wall but almost everywhere in the boundary layer. They can be observed by optical means such as schlieren photographs (Roshko, 1976) particle imaging velocimetry (PIV) or defined by measurements of coherency between velocity signals separated in time and space e.g. (Antonia, 1980). The number, size and shape of these so-called coherent structures found so far is bewildering although one can recognise linkages of the patterns to vorticity, flow geometry and kinetic/dissipation energy ratios. Cantwell (1981) and Robinson (1991) have reviewed coherent structures in detail.

Oblique shear layers have been observed near the wall and upstream of large scale structures by Blackwelder and Kovasnay (1972), Nychas et al. (1973), Hedley and Keffer (1974), Falco (1977), Brown & Thomas (1977), Chen & Blackwelder (1978), Spina and Smits (1987), and Antonia et al. (1989). It is not clear from literature reports whether these observations refer exactly to the same phenomenon and what effects the different methods of event detection have on the results. There is, however, a consensus that a secondary structure exists, probably created by ejections from the wall. One interesting visualisation was proposed by Townsend (1970) and by Grass (1971) who viewed the ejections from the wall as similar to intermittent jets of fluid essential at cross-flow with the main stream. The discovery of coherent structures implies the existence of strong secondary currents embedded in the turbulent field. One would expect that the relative motion between these currents and the main flow would create intense shear layers. The existence of shear layers can be monitored by estimating the contribution of transient $\langle u'v' \rangle$ peaks to the total local shear stress. Johansson et al. (1991) analyzed the DNS data base of Kim et al. (1987) and confirmed that the Reynolds stresses contribution form the downstream side of the shear layers is spatially spotty but they could follow the associated $\langle u'v' \rangle$ peaks for distances up to 1000 wall units (equal to $v/u_*$).

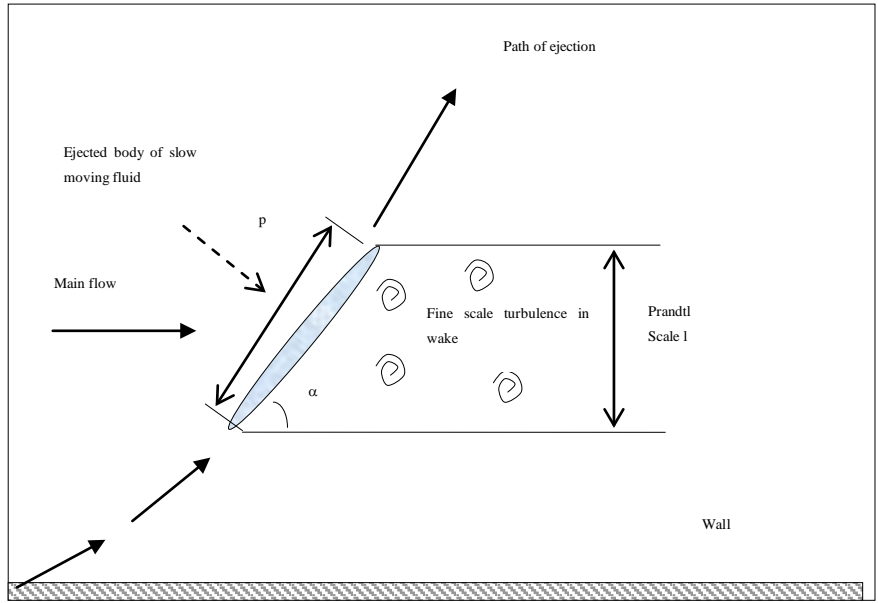

Figure 2. Schematic representation of ejection as a intermittent jet from the wall

Furthermore, they found no sign of oscillatory motions or violent break-up in conjunction with these shear layers which, they believe, indicate a persistent motion of low-speed fluid away from the wall. A schematic representation of these observations is shown in Figure 2.

Thus it appears that the typical scale of turbulence may be more indicative of the size of a slowly recirculating body of fluid which is strongly ejected into the mainstream, the eddy scale interpretation of Taylor (1935) rather than the mean free path of this fluid body before it disintegrates, the Prandtl mixing-length postulate. Upon interaction with this ejected body of fluid, the mainstream flow breaks up and produces small scale turbulence but the ejected fluid body itself persists.

## 2.2 The velocity profile

The local time-averaged shear stress at a distance y is given by

$$\tau = \tau_v + \tau_t = \tau_v + \overline{U'V'} \qquad (2)$$

where  $\tau_v$  is  the time-averaged viscous shear contribution, and

$\tau_t$   the turbulent or eddy shear stress contribution

$U', V'$  the fluctuating velocities in the x and y directions.

And can be written as

$$\tau = (\mu + \rho E_\nu)\frac{dU}{dy} \tag{3}$$

where $E_\nu$ is called the eddy viscosity after a proposal by Boussinesq (1877). Rearranging in dimensionless form gives

$$U^+ = \int_0^{y^+} \frac{\tau/\tau_w}{1 + E_\nu/\nu} dy^+ \tag{4}$$

Prandtl assumed that the fluctuating velocity can be obtained by expanding the velocity into a Taylor series and keeping only the first term

$$U' \approx V' \approx \Delta U \approx l\frac{dU}{dy} \tag{5}$$

and the turbulent Reynolds stresses and the eddy viscosity as

$$\tau_t = \overline{U'V'} = \rho l^2 \left(\frac{dU}{dy}\right)^2 = \rho E_\nu \frac{dU}{dy} \tag{6}$$

Hence

$$E_\nu = l^2 \frac{dU}{dy} = l\sqrt{\frac{\tau_t}{\rho}} \tag{7}$$

Equation (4) may be rearranged as

$$\frac{dU^+}{dy^+} = \frac{\tau/\tau_w}{1 + E_\nu/\nu} \tag{8}$$

In the log law region, the viscous contribution is small and $1 \ll E_\nu/\nu$ and we may write

$$\frac{dU^+}{dy^+} \cong \frac{\tau/\tau_w}{E_\nu/\nu} = \frac{\tau/\tau_w}{(l/\nu)\sqrt{\tau_t/\rho}} = \frac{\tau/\tau_w}{(lu_*/\nu)\sqrt{\tau_t/\tau}\sqrt{\tau/\tau_w}} \tag{9}$$

Prandtl assumed that

$$l = \kappa y \tag{10}$$

Substituting for equation (10)

$$\frac{dU^+}{dy^+} = \frac{\sqrt{\tau/(\tau_w \tau_t)}}{\kappa y^+} \tag{11}$$

which does not lead to a log-law. To obtain the log law from equation (11) Prandtl had to assume further that the shear stress in the log-law layer is uniform

$$\tau = \tau_w = \tau_t \tag{12}$$

In the log-law layer, the viscous shear stress is assumed to be negligible Following Prandtl, we can express the velocity gradient as

$$\frac{dU^+}{dy^+} = \frac{1}{\kappa y^+} \tag{13}$$

and

$$U^+ = \frac{1}{\kappa} \ln y^+ + B \tag{14}$$

Near the wall, Prandtl's logarithmic law breaks down and a damping function must be included in the formulation

$$l = \kappa y F \tag{15}$$

The oldest as that of van Driest (1956).

$$F = \left(1 - e^{-0.03 y^{+2}}\right) \tag{16}$$

## 2.3 Significance of Karman's universal constant

A Taylor expansion only makes sense of course if the velocity fluctuations in the domain of application are correlated. In the writer's view, the size p of the body of fluid ejected from the wall can be interpreted as an eddy scale since the velocity fluctuations within that body will be strongly correlated even while it is moving through the main flow field. However, since the transient shear layers created between the ejections and the main flow are inclined at an angle $\alpha$ with respect to the wall, it appears that these turbulent shear stresses are more adequately estimated by using coordinates which coincide with the path of the ejections rather than by using the normal distance $y^+$ proposed by Prandtl.

The writer suggests that Karman's universal constant in equation (1) is the sine of the angle of projection α as sketched in Figure 2. The Prandtl scale l is only a projection of the scale p of the shear layers onto the y axis.

$$\kappa = \sin \alpha = \frac{l}{p} \tag{17}$$

A similar situation exists in the analysis of jets in cross flow. It is customary to investigate the progression of the jet using "natural" coordinates based on the jet geometry as shown in Figure 3. The axis y' is the centre stream-line of the jet flow. A similar representation of jet penetration is achieved using the natural coordinates but no similarity pattern is found using Cartesian coordinates. While our understanding of jets in crossflow is still very incomplete, it appears that most of the parameters, for example the distribution of lateral velocity of the jet are best correlated in terms of the natural coordinates (Chan *et al*, 1976; Keffer & Baines 1963).

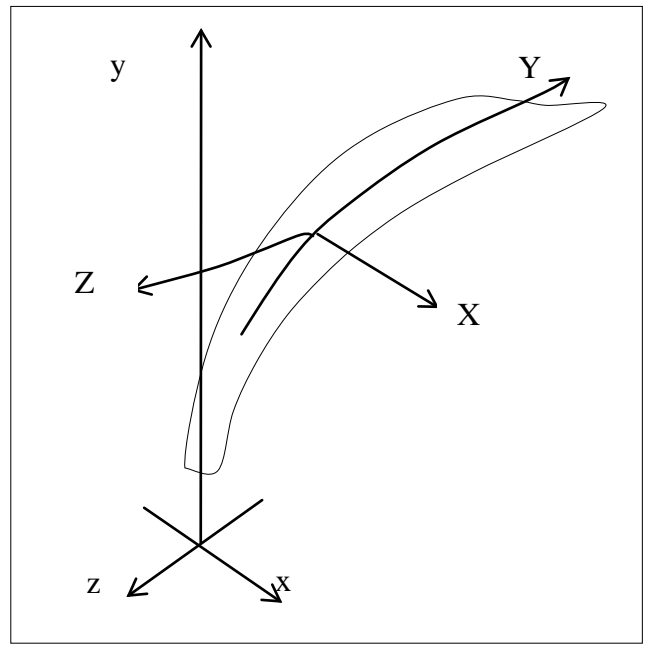

Figure 3. Natural and Cartesian coordinates for a jet in cross flow

### 2.4  Estimate of the inclination of shear layers

In the log-law region, where the viscous contribution to the local shear stress can be neglected, the coefficient A can be expressed in terms of κ by combining equations (1) and (7). Then

$$\frac{1}{A} = \kappa \left( \frac{\tau}{\tau_w} \right) \qquad (18)$$

and the angle α can be estimated from equation (18). Near the wall, Prandtl's logarithmic law breaks down and a damping function must be included in the formulation such as that of van Driest (1956) or Deissler (1955). We use here the formulation of Trinh (1992) which is similar to that of Khishinevskii and Kornienko (1967)

$$\frac{1}{A} = \kappa \left(\frac{\tau}{\tau_w}\right)\left(1 - e^{-0.03 y^{+2}}\right) \qquad (19)$$

and the angle α is again estimated from equation (17) and (19).

## 3  Comparison with experimental data

### 3.1  Measurements of coherent structures by correlation techniques by Brown and Thomas and Kreplin and Eckelmann

Using an array of hot wires and wall shear stress probes, Brown & Thomas (1977) made correlations of the wall shear and the local instantaneous velocity. The long time correlations exhibited clear peaks which could be detected at positions well away from the wall, inside the turbulent flow stream. When the probes were placed at different normal distances but aligned on the same x plane, the peak of the correlations shifted with time as shown in Figure 4a. The x position of the correlation peak was deduced by multiplying the time shift of the peak with the local time-averaged convection velocity. The probes were then staggered in the x direction by trial and error to obtain all peaks at t=0. The positions of the wires were found to match very well with the values calculated by the previous method. Furthermore, they coincided with the upstream side of large structures observed by flow visualisation techniques (Figure 4b).

Brown and Thomas interpreted these measurements as indicative of the existence of an organised structure slanted at an oblique angle with respect to the wall and related in some sense to the bursting process at the wall.

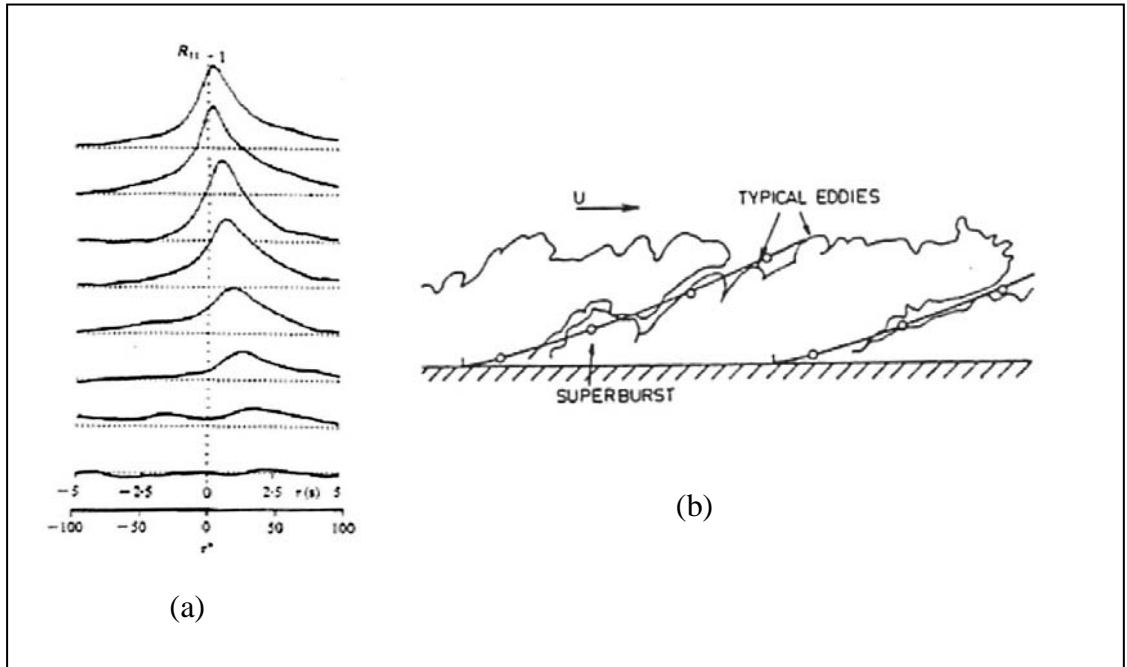

Figure 4 (a) Shift in correlation peak (b) Positions of correlation peaks and coherent structure. After Brown and Thomas (1977)

Kreplin & Eckelmann (1979) also made similar wall shear stress-velocity correlations within the wall layer. They transformed the time shift in the peaks of their correlations into spacing in the x direction by multiplying with a convection velocity measured by wall probes spaced in the x direction and called the educed trace "a moving front of turbulence" in the wall layer.

### 3.2 Comparison between calculated and measured traces of the shear layers

At each position $y^+$, the mixing length l and the coefficient A were calculated simultaneously from experimental measurements of the time-averaged velocity and shear profiles in Nikuradse's work, and the angle α was calculated. Since the position $x^+$ of the shear layer p is not known but the value of $y^+$ is, the picture is created by dovetailing the angle of inclination of successive vectors.

In the wall layer, the resulting trace coincides well with the 'front of turbulence' obtained by Kreplin and Eckelmann as shown in Figure 5.

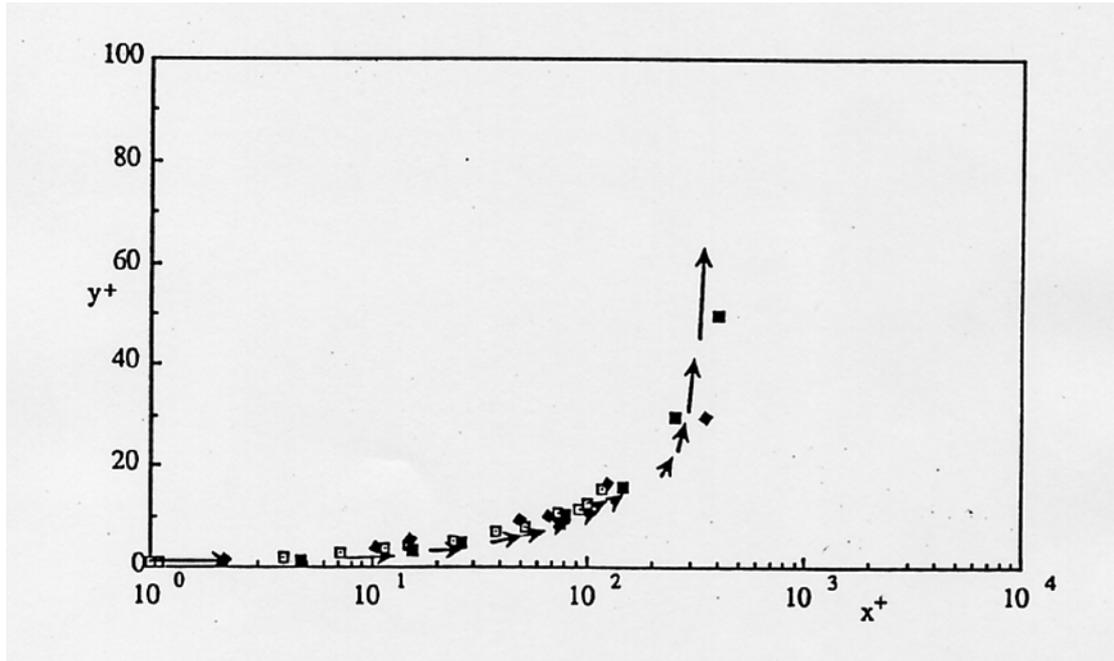

Figure 5 Comparison between predicted inclination of shear layer and front of turbulence measured by Kreplin and Eckelmann (1977)

Outside the wall layer, it coincides with the inclined front deduced by Brown and Thomas (1977) from measurements of correlation peaks as shown in Figure 6.

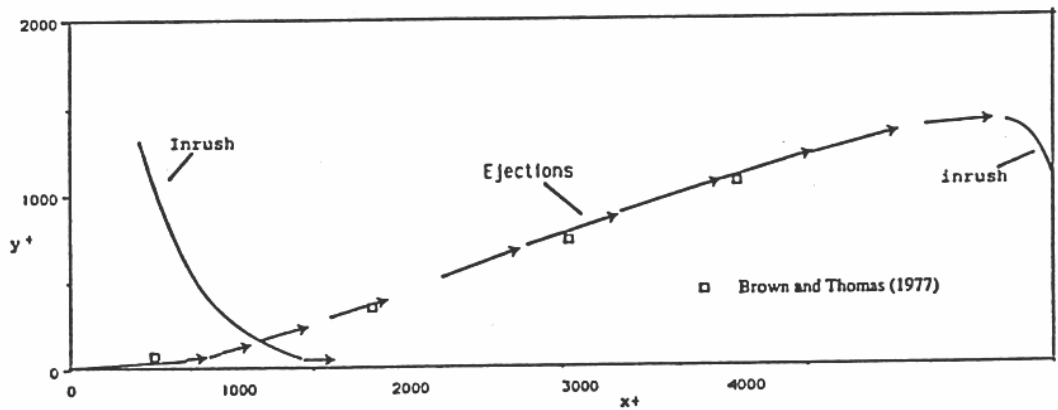

Figure 6 Position of correlation peaks (Brown and Thomas, 1977) and predicted trace

## 4  Discussion and conclusion

The excellent agreement between predictions and measurement is encouraging and indicates that the Karman universal constant can be interpreted as a geometric structural parameter, at least in the log-law region. It appears that the estimate of turbulent stresses in a flow field is best made by correlations along their angle of inclination rather than by using the normal coordinate y. However, since this angle of inclination varies continuously as one move away from the wall, and possibly with flow geometry, more investigation concerning the inclination of the shear layers is required before practical applications are possible. It is most likely that the inclination of the shear layer will depend on the relative strengths of the main flow and the ejection. It is therefore remarkable that Karman's constant is confined to a narrow range around the canonical value of 0.4. One possible explanation is that ejections always occur when the ratio of kinetic to viscous energy in the transient sub-boundary layer reaches a critical value as indicated by the estimate of the instantaneous friction factor at the end of the sweep phase when bursting occurs (Trinh, 2009).

The present paper has shown only that the trace of moving peaks in correlations between the wall shear stress and the instantaneous velocity can be obtained from measurements of time-averaged shear stresses and local velocities, through an estimate of the mixing-length and Karman's universal constant. As mentioned earlier, Brown & Thomas have shown that this trace can be overlaid perfectly on smoke photographs of a boundary layer. They interpret this result to indicate that a large inclined coherent structure exists in the turbulent flow field. On the other hand, Kreplin & Eckelmann interpret their measurements as the trace of a moving "front of turbulence" near the wall. Johansson *et al* interpret similar data as indicative of the motion of patches of slow-peed fluid ejected from the wall.

Within existing evidence, it is not possible to conclusively differentiate between these postulates although they may be reconciled as parts of the jig-saw puzzle. The ambiguity with which statistical probe correlations can be interpreted in physical terms remains one of the major difficulties in turbulence research.